\documentclass[aps,amsfonts,nofootinbib,article,twocolumn,showpacs,superscriptaddress]{revtex4}
\usepackage{graphicx}
\usepackage{amsmath}
\usepackage{amsfonts}
\usepackage{amssymb}

\def\openone{\leavevmode\hbox{\small1\kern-3.8pt\normalsize1}}
\def\N{\leavevmode\hbox{ Z \kern-8 pt\normalsize{Z}}}
\def\openone{\leavevmode\hbox{\small1\kern-3.8pt\normalsize1}}
\def\openJ{\leavevmode\hbox{J \kern-9.5pt\normalsize J}}
\def\openS{\leavevmode\hbox{ S \kern-9.3pt\normalsize S}}
\newcommand{\bb}{\begin{equation}}
\newcommand{\ee}{\end{equation}}
\newcommand{\eqb}{\begin{eqnarray}}
\newcommand{\eqf}{\end{eqnarray}}

\usepackage{color}

\begin{document}

\title{Class of exact solutions for a cosmological model of unified gravitational and quintessence fields}

\author{Felipe A. Asenjo}
\email{felipe.asenjo@uai.cl}
\affiliation{Facultad de Ingenier\'{\i}a y Ciencias,
Universidad Adolfo Ib\'a\~nez, Santiago, Chile.}
\author{Sergio A. Hojman}
\email{sergio.hojman@uai.cl}
\affiliation{Departamento de Ciencias, Facultad de Artes Liberales,
Universidad Adolfo Ib\'a\~nez, Santiago, Chile.}
\affiliation{Facultad de Ingenier\'{\i}a y Ciencias, Universidad Adolfo Ib\'a\~nez, Santiago, Chile.}
\affiliation{Departamento de F\'{\i}sica, Facultad de Ciencias, Universidad de Chile,
Santiago, Chile.}
\affiliation{Centro de Recursos Educativos Avanzados,
CREA, Santiago, Chile.}

\begin{abstract}
A new approach to tackle Einstein equations for an isotropic and homogeneous  Friedmann--Robertson--Walker Universe in the presence of a quintessence scalar field is devised. It provides a way to get a simple exact solution to these equations. This solution determines the quintessence potential uniquely and it differs from solutions which have been used to study inflation previously. It relays on a unification of geometry and dark matter implemented through the definition of a functional relation between the scale factor of the Universe and the quintessence field. For a positive curvature Universe, this solution produces perpetual accelerated expansion rate of the Universe, while the Hubble parameter increases abruptly, attains a maximum value and decreases thereafter. The behavior of this cosmological solution is discussed and its main features are displayed. The formalism is extended to include matter and radiation.
\end{abstract}

\pacs{}

\maketitle

\section{introduction}

One way to study the effect of dark energy  in a cosmological model is by describing it by means of a scalar field with a potential energy. This field, sometimes called quintessence field or phantom scalar field, has been widely explored in inflationary theory (the literature is vast, see for example Refs.~\cite{ryden, tsuji, gutha,liddle,watson,brande,linde}). The equations which describe the model are usually solved approximately by considering  that the potential energy of the field is always much larger than its kinetic energy, producing inflation.

Our approach is based on a simple and interesting  feature of the Friedmann--Robertson--Walker--Quintessence (FRWQ) equations. A direct identification between the scale factor of the Universe and the quintessence field allows for a unique determination of the quintessence potential and produces equations that may be solved exactly. The solution shows an accelerated expansion of the Universe different from other inflationary theories.  It can be also proved that this solutions admits the inclusion of matter or radiation in a consistent fashion.

First we start studying the solution for quintessence field highlighting the main features of this approach. Then, in Sec. III, we include a component to the Universe showing that the same anstaz allow us to solve the cosmological equations.

To write the equations for the FRWQ model, we start with the Lagrangian density $\mathcal{L}$ for the evolution of the
spacetime metric $g_{\mu \nu}(x^\alpha)$ in interaction with a
 scalar  field $\phi(x^\beta)$. This is $\mathcal{L}=\sqrt{-g}\left[(R-2\Lambda)/2{\cal G}+{\mathcal L}_\phi\right]$,
where $g$ stands for the determinant of the metric $g_{\mu\nu}$, ${\cal G}= 8\pi G/c^4$ (with $G$ as the gravitational constant and $c$ the speed of light) and $\Lambda$ is the  cosmological constant. Here, $\mathcal{L}_\phi$ is the Lagrangian density for the  scalar field
\begin{equation}
\mathcal{L}_\phi=\epsilon\left(\frac{1}{2}g^{\mu \nu} \phi,_\mu
\phi,_\nu - {\cal V}(\phi)\right)\, ,\label{lfrwq2}
\end{equation}
where ${\cal V}(\phi)$ is  (up to now) an unspecified potential for the
scalar field $\phi$, and $\epsilon$ is a parameter that classifies the nature of the scalar field, i.e.,  $\epsilon=1$ yields the Lagrangian density for usual scalar fields while $\epsilon=-1$ defines the  Lagrangian for a quintessence field \cite{Steinhardt}.
The line element for an isotropic and
homogeneous Friedmann--Robertson--Walker (FRW) spacetime is \cite{ryden}
\begin{equation}
ds^2=dt^2- a(t)^2\left[\frac{d r^2}{1-k r^2} +r^2(d\theta^2+
\sin^2\theta d\phi^2)\right]\, , \label{frwm}
\end{equation}
that defines the metric, where $a(t)$ is  the scale factor of the Universe, and  $k$ is  the curvature constant that takes the values $k=-1, 0, 1$  denoting negative, zero or positive curvature of the Universe.
Furthermore, it is worth noting that the quintessence field is described by the same Lagrangian density as the tlaplon field \cite{hrrs} which describes propagating torsion.

The action $S= \int \mathcal{L}\ d^4 x \label{action}$,
gives rise (upon variation with respect to the metric tensor $g_{\mu
\nu}$) to gauge invariant (generally covariant) and constrained
 Einstein field equations coupled to matter $G_{\mu \nu}+ \Lambda g_{\mu\nu} ={\cal G}\, T_{\mu \nu}$,
where $G^{\mu \nu}$ is the Einstein tensor and $T^{\mu \nu}$ is the
energy--momentum tensor of matter $T_{\mu\nu}=g_{\mu\nu} \mathcal {L}_\phi-2\, {\partial \mathcal {L}_\phi}/{\partial g^{\mu\nu}}$.
Assuming that the scalar field $\phi$ depends on time only, the Einstein equations, in terms of
the line element \eqref{frwm}, get two second order dynamical equations and one first order constraint. The dynamical equations read (setting ${\cal G}=1$)
\begin{equation}
2\frac{\ddot a}{a}+\left(\frac{\dot a}{a}\right)^2 +\
\frac{k}{a^2} -\epsilon\left(\frac{1}{2}\ \dot \phi^2-V(\phi)
\right)= 0, \label{frwqe1}
\end{equation}
and
\begin{equation}
\ddot\phi + 3\frac{\dot a}{a} \dot\phi + \frac{d
V(\phi)}{d\phi}=0\, , \label{frwqe2}
\end{equation}
where we have introduced $V(\phi)={\cal V}(\phi)+\epsilon \Lambda$. The constraint equation is
\begin{equation}
3\left(\frac{\dot a}{a}\right)^2 +3\frac{k}{a^2}+\epsilon\left(\frac{1}{2}\dot \phi^2+ V(\phi) \right) =0\, . \label{frwqc}
\end{equation}
Eq.~\eqref{frwqe2} is obtained by varying the action $S$ with respect to $\phi$, producing the Klein--Gordon equation for the  scalar field $\Box \phi + {d {\cal V}(\phi)}/{d \phi}= 0$.
On the other hand, another useful equation can be obtained manipulating Eqs. \eqref{frwqe1} and \eqref{frwqc} to give $3{\ddot a}/{a}={\epsilon}\left(\dot\phi^2-V\right)$.

For $\epsilon=-1$, the set \eqref{frwqe1}-\eqref{frwqc} becomes the  FRWQ system. This set of equations has been already studied   for different scenarios and matter configurations \cite{tsu,copeland,capozz}.
The FRWQ system is usually employed to study inflation under standard approximations $\dot\phi^2\ll V(\phi)$ and $\ddot\phi \ll dV/d\phi$ \cite{ryden}. That solution represents exponential (inflationary) expansion of the Universe.

\section{Exact solution}

The purpose of this manuscript is to show that there is an exact solution that may be obtained by defining a specific relation between the space--time curvature (represented by the scale factor $a$) and dark energy (represented by the quintessence field $\phi$).  As the scalar field depends on time, then $\phi\equiv\phi(a)$,  $V= V(\phi)\equiv V(a)$, and $\dot\phi=({d \phi}/{d a})\dot a$. Therefore, Eq.~\eqref{frwqc} becomes
\begin{equation}\label{eqconsSolved}
\dot a^2 \left[\frac{3}{a^2} +\frac{\epsilon}{2}\left(\frac{d \phi}{d a}\right)^2 \right]=-\epsilon V(a)-3\frac{k}{a^2}\, .
\end{equation}

Notice that for $\epsilon=-1$ a non--trivial solution can be obtained if the left-hand side of the previous equation is forced to vanish.
With this choice, the quintessence scalar field $\phi$ is related to the scale factor $a$ by
\begin{equation}\label{unificaVariable}
\phi=\sqrt{6}\ln a\, ,
\end{equation}
and it determines the quintessence potential as
\begin{equation}\label{QPot}
V=\frac{3k}{a^2}\, .
\end{equation}
This kind of ansatz  cannot be performed for $\epsilon=1$.
 It is worth to mention that other attemps
of relating the quintessence field to the metric have been proposed \cite{AliH}.

It is a straightforward matter to show that the solutions given by Eqs.\eqref{unificaVariable}-\eqref{QPot}, both Eqs.~\eqref{frwqe1} and \eqref{frwqe2} give rise to the same following equation
\begin{equation}\label{unifiedGQ}
\frac{\ddot a}{a}+2\left(\frac{\dot a}{a}\right)^2-\frac{k}{a^2}=0\, ,
\end{equation}
in terms of $a$ (or an equivalent equation written in terms of $\phi$). Remarkably,  Eq.~\eqref{unifiedGQ} can be solved exactly, describing a cosmological model where gravity and dark energy are intimately intertwined into a single structure \eqref{unificaVariable}.
This  result is due to the fact that \eqref{frwqc} is, loosely speaking, the ``energy conservation equation'' derived from the dynamical equations ~\eqref{frwqe1} and \eqref{frwqe2}.

We can infer directly from Eq.~\eqref{unifiedGQ} that this cosmological model presents accelerated expansion $\ddot a>0$ for $k=1$ if the (positive) initial expansion velocity ${\dot a}(0)$ is restricted to be
\begin{equation}\label{condforvelocity}
0<{\dot a}(t=0)\ <\ \frac{1}{\sqrt 2}\, ,
\end{equation}
being a particular initial condition of the system.  Notice that in this case, the quintessence scalar field becomes a dilaton field with the potential
\begin{equation}\label{potenExp}
V(\phi)=3\exp\left(-\frac{2\phi}{\sqrt 6}\right)\, .
\end{equation}
For the other cases, either $k=0$ or $k=-1$, this cosmological model does not produce accelerated expansion. Although with different cosmological models with respect to the one presented in this work, several potentials with the form of \eqref{potenExp} have been considered extensivelly in the literature \cite{salopek,neupane,ruso,elizalde,fre2}, showing that these models are integrable even in the case when the Hubble parameter is written in terms of the quintessence scalar field \cite{salopek,musilov,bazeia,harko}.

On the other hand, Eq.~\eqref{unifiedGQ} can be integrated to find a conservation law. This result to be
\begin{equation}\label{unifiedGQ2}
E=a^4\left(\frac{k}{2}-\dot a^2\right)\, ,
\end{equation}
for general $k$, and where $E$ is a constant. In the case $E=0$, then the dynamics is restricted to $\ddot a=0$. We can see  that for $k=1$ and under the initial condition \eqref{condforvelocity}, the constant $E$ is positive.
The previous equation can be integrated to find the general solution of Eq.~\eqref{unifiedGQ}
\begin{equation}\label{solutya}
t=\int_0^a \frac{a^2da}{\left(k a^4/2-E\right)^{1/2}}\, ,
\end{equation}
can be exactly written in terms of elliptical integrals for $k=1$.
As an aside, notice that defining $u=a^3/3$, Eq.~\eqref{unifiedGQ} becomes $\ddot u=k \left(3 u\right)^{1/3}$, showing that this kind of cosmology behaves  as a particle driven by to the force $\left(3 u\right)^{1/3}$ for a  Universe with positive curvature.

Let us study the behavior of the solutions of $\dot a$, $\ddot a$ and the Hubble parameter $H=\dot a/a$ for certain initial conditions $a(t=0)\ll 1$ and $\dot a(t=0)\ll1$.
Let us start the analysis for $k=1$. First, we notice that from Eq.~\eqref{unifiedGQ2} the initial conditions of $a$ and $\dot a$ (or $H$) determine $E$. As initially $\dot a$ is small, then from Eq.~\eqref{unifiedGQ2} we get that $0<E\ll 1$. Also, from Eq.~\eqref{solutya} we see that our solution has the lower limit $a>(2E)^{1/4}$.
From the evolution equation \eqref{unifiedGQ}, we see that if the expansion is accelerated $\ddot a\geq 0$,
then $\dot a$ grows with time reaching the  value $1/\sqrt 2$ asymptotically, where the accelerated expansion ends. Close to $1/\sqrt 2$, the scale factor grows linearly with time. Thereby, in previous times, the time derivative of the scale factor is always positive. The exact behavior of the expansion is obtained from \eqref{unifiedGQ2}
\begin{equation}\label{solucionaterma}
\dot a(t)=\sqrt{\frac{1}{2}-\frac{E}{a(t)^4}}\, ,
\end{equation}
where the time-dependence of $a$ is obtained from Eq.~\eqref{solutya}.
The behavior of $\ddot a$ can be obtained from Eqs.~\eqref{unifiedGQ} and \eqref{unifiedGQ2}. As long as $k=1$ and  $\dot a(0)< 1/\sqrt 2$, the expansion is accelerated $\ddot a>0$. Using Eqs.~\eqref{unifiedGQ} and \eqref{unifiedGQ2}, we  can obtain that
\begin{equation}\label{addota5}
\ddot a=\frac{2E}{a^5}\, ,
\end{equation}
and the initial condition for $\ddot a(0)$ is determined by $a(0)$. Solution \eqref{addota5} shows that although in this model the Universe is subjected to an accelerated expansion, the acceleration decreases as $a^{-5}$.
Finally, we can study the Hubble parameter, obtaining from Eq.~\eqref{solucionaterma} that
\begin{equation}\label{solucionaterma2}
H=\sqrt{\frac{1}{2a^2}-\frac{E}{a^6}}\, .
\end{equation}
As $\dot a >0$, then the Hubble parameter grows reaching the maximum value  $(54\ E)^{-1/4}$ for $a=(6E)^{1/4}$. In this point, $\dot a=1/ \sqrt 3$ and $\ddot a=(486\ E)^{-1/4}$. Then, it decreases with time as $\dot a$ approaches $1/\sqrt{2}$.
An approximatted solution of Eqs.~\eqref{unifiedGQ} and \eqref{unifiedGQ2} can be obtained using the above information. As $E\ll 1$, it is starightforward to obtain that
\begin{equation}
a=\frac{t}{\sqrt 2}+\frac{2\sqrt{2}E}{3t^3}+\mathcal{O}(E^2)\, .
\end{equation}
This shows that the acceleration of the expansion for $k=1$ goes as
\begin{equation}
\ddot a\approx\frac{8\sqrt{2} E}{t^5}>0\, .
\end{equation}

On the other hand, similar analisis can be performed to the $k=0$ case. From  Eqs.~\eqref{unifiedGQ} and \eqref{unifiedGQ2} we infer that $E<0$. In this case we can direct integrate to obtain
\begin{equation}
a=\left(3\sqrt{-E}\, t\right)^{1/3}\, ,
\end{equation}
producing a desaccelerating expansion rate given by
\begin{equation}
\ddot a=-\frac{2\left(3\sqrt{-E}\right)^{1/3}}{9 t^{5/3}}\, .
\end{equation}

Lastly, for the $k=-1$ model, we get from  Eqs.~\eqref{unifiedGQ} and \eqref{unifiedGQ2} that $E<0$. Eq.~ \eqref{unifiedGQ2} also gives the upper limit to the dynamics  $(-2E)^{1/4}>a$. As $E$ is small by  initial conditions, the dynamics of this case is not very interesting.

\section{Model of Universe with different components}

One may wonder if this kind of solution holds for a Universe with single components. For these cases, the Eqs.~\eqref{frwqe1}-\eqref{frwqc} must be extended to include the effect of matter or radiation. The pertinent equations for such purposes are
\begin{equation}
2\frac{\ddot a}{a}+\left(\frac{\dot a}{a}\right)^2 +\
\frac{k}{a^2} -\epsilon\left(\frac{1}{2}\ \dot \phi^2-V(\phi)
\right)= -p, \label{frwqe1b}
\end{equation}
\begin{equation}
3\left(\frac{\dot a}{a}\right)^2 +3\frac{k}{a^2}+\epsilon\left(\frac{1}{2}\dot \phi^2+ V(\phi) \right) = {\bf \varepsilon}\, , \label{frwqcb}
\end{equation}
\begin{equation}
\ddot\phi + 3\frac{\dot a}{a} \dot\phi + \frac{d
V(\phi)}{d\phi}=0\, , \label{frwqe2b}
\end{equation}
and
\begin{equation}
\dot\varepsilon + 3\frac{\dot a}{a}(\varepsilon+p)=0\, , \label{fluidb}
\end{equation}
where $\varepsilon$ is the energy density of the corresponding component, whereas $p$ is its pressure. The whole previous system is consistent, and  the closure is given by the equation of state $p=w \varepsilon$, with $w$ a dimensionless number. For matter-dominated era of the Universe $w=0$, while for the radiation-dominated era, $w=1/3$.

Newly, the system can be written in unified form for the gravitational and quintessence fields if we take the solution \eqref{unificaVariable} [$\epsilon=-1$ and $\phi=\sqrt{6}\ln a$]. In this way, the quintessence potential acquires the form
\begin{equation}\label{QPotb}
V=\frac{3k}{a^2}-\varepsilon\, .
\end{equation}
This ansatz solves identically  Eq.~\eqref{frwqcb}. Besides, Eqs.~\eqref{frwqe1b} and \eqref{frwqe2b} become the following equation
\begin{equation}\label{unifiedGQb}
\frac{\ddot a}{a}+2\left(\frac{\dot a}{a}\right)^2-\frac{k}{a^2}+\frac{\varepsilon}{2}(1+w)=0\, ,
\end{equation}
where we have made use of Eq.~\eqref{fluidb}
\begin{equation}
\dot\varepsilon + 3\frac{\dot a}{a}\varepsilon(1+w)=0\, . \label{fluidc}
\end{equation}

As the solution of \eqref{fluidc} is
\begin{equation}
\varepsilon=\varepsilon_0 a^{-3(1+w)}\, ,
\end{equation}
with the initial value of $\varepsilon_0$, then this FRWQ model with single component is determined by the equation
\begin{equation}\label{unifiedGQc}
\frac{\ddot a}{a}+2\left(\frac{\dot a}{a}\right)^2-\frac{k}{a^2}+\frac{\varepsilon_0(1+w)}{2 a^{3(1+w)}}=0\, .
\end{equation}
The previous dynamics can be integrated for $w\neq 1$ to give
\begin{equation}\label{unifiedGQ2b}
E=a^4\left[\frac{k}{2}-\dot a^2-\frac{\varepsilon_0}{3 a^{1+3w}}\left(\frac{1+w}{1-w}\right)\right]\, ,
\end{equation}
where $E$ is again a constant. The solution can be obtained through the integral (which can be  solved in terms of elliptical integrals)
\begin{equation}\label{integralmatterradiua}
t=\int da\left[\frac{k}{2}-\frac{E}{a^4}-\frac{\varepsilon_0}{3a^{1+3w}}\left(\frac{1+w}{1-w}\right)\right]^{-1/2}\, ,
\end{equation}
for matter-dominated Universe ($w=0$) and  radiation-dominated Universe ($w=1/3$).

Again we can study the behavior of the solutions  for initial conditions $a(t=0)\ll 1$ and $\dot a(t=0)\ll1$.
From Eq.~\eqref{unifiedGQc} we infer that $E\ll 1$ for $w=0$ and $w=1/3$.
We can readily see that the only case that produce accelerated expansion $\ddot a>0$ is for $k=1$. For this case is enough to choose $E>0$, and the dynamics is restricted to a lowest limit given by
\begin{equation}
a\left[1-\frac{2\varepsilon_0}{3 a^{1+3w}}\left(\frac{1+w}{1-w}\right)\right]>(2E)^{1/4}\, .
\end{equation}
Similar to the previous case, as $E\ll 1$, an approximated solution can be found $a(t)=a_1(t) +E\, a_2(t)+\mathcal{O}(E^2)$, where $a_1$ can always be found as the solution of integral \eqref{integralmatterradiua} with $k=1$ and $E=0$, whereas $a_2$ satifies the equation
\begin{equation}
\left(a_1^4\dot a_1\right)\dot a_2=\frac{\varepsilon_0 a_2}{2a_1^{-2+3w}}\left(\frac{1+w}{1-w}\right)\left(\frac{1}{3}+w\right)-\frac{1}{2}\, .
\end{equation}

For the $k=0$ and $k=-1$ case the dynamics do not present accelerated expansion. Also $E<0$, and  they both are restricted to an upper limit.

\section{Conclusions}

We have shown that the previous solutions produce a  functional relation between the Universe scale factor $a$ and the dark matter quintessence field, uniquely determining the scalar field $\phi = \phi (a)$ [Eq.~\eqref{unificaVariable}], and  the quintessence potential [Eqs.~\eqref{QPot} and \eqref{QPotb}]. This approach constitutes a new solution of the FRWQ system, as implies a new ansatz that can produce a new dynamic different to previous ones where the scalar field and the potential are reconstructed from the equations of motion \cite{manti}. This solution is different to previous models as the relation between spacetime and the fields is a priori invoked to solve the dynamics.

For the case of only quintessence fields, it can be easily seen that the evolution of the scale factor $a$ for $k=1$ is such that its velocity reaches asymptotically the value $1/\sqrt 2$. Therefore, the evolution is always positively accelerated,  while the Hubble parameter grows until it reaches a maximum value and it decreases from then on. This behavior is appropiated for an initial value of $\dot a$ constrained to be less than $1/\sqrt 2$. In this case, the Universe attains the terminal velocity  $1/ \sqrt{2}$, at $a\rightarrow \infty$ behaving as an attractor point for the Universe.
If $\dot a$ is larger than $1/\sqrt 2$ initially, then the Universe experiences a decelerated expansion with $\dot a$ also approaching to $1/ \sqrt{2}$ at the attractor point $(a \rightarrow \infty,\  {\dot a} \rightarrow  1/ \sqrt{2})$.
On the contrary,  we can see from \eqref{unifiedGQ} that for $k\neq 1$ the Universe always experiences a decelerated expansion $\ddot a<0$.

Also, we have shown that the ansatz \eqref{unificaVariable} and \eqref{QPot} still holds when matter or radiation are included in the scheme. Again the most interesting dynamics is achieved for $k=1$ showing that the system can be solved for $E\ll1$.

\begin{acknowledgments}
F.A.A. thanks the CONICyT-Chile for partial support through Funding No. 79130002. S.A.H. expresses his gratitude to Rafael Rosende for his enthusiastic support.
\end{acknowledgments}

\end{document}